\documentclass[aps, prl, showpacs, superscriptaddress, twocolumn]{revtex4}
\usepackage{graphicx}
\usepackage{hyperref}

\begin{document}

\title{Layer-resolved study of Mg atom incorporation at MgO/Ag(001) buried interface}

\author{T. Jaouen}
\altaffiliation{Corresponding author.\\ thomas.jaouen@unifr.ch}
\affiliation{D{\'e}partement de Physique and Fribourg Center for Nanomaterials, Universit\'e de Fribourg, CH-1700 Fribourg, Switzerland}
\affiliation{D{\'e}partement Mat{\'e}riaux Nanosciences, Institut de Physique de Rennes UMR UR1-CNRS 6251, Universit{\'e} de Rennes 1, F-35042 Rennes Cedex, France}

\author{S. Tricot}
\affiliation{D{\'e}partement Mat{\'e}riaux Nanosciences, Institut de Physique de Rennes UMR UR1-CNRS 6251, Universit{\'e} de Rennes 1, F-35042 Rennes Cedex, France}

\author{G. Delhaye}
\affiliation{D{\'e}partement Mat{\'e}riaux Nanosciences, Institut de Physique de Rennes UMR UR1-CNRS 6251, Universit{\'e} de Rennes 1, F-35042 Rennes Cedex, France}

\author{B. L{\'e}pine}
\affiliation{D{\'e}partement Mat{\'e}riaux Nanosciences, Institut de Physique de Rennes UMR UR1-CNRS 6251, Universit{\'e} de Rennes 1, F-35042 Rennes Cedex, France}

\author{D. S{\'e}billeau}
\affiliation{D{\'e}partement Mat{\'e}riaux Nanosciences, Institut de Physique de Rennes UMR UR1-CNRS 6251, Universit{\'e} de Rennes 1, F-35042 Rennes Cedex, France}

\author{G. J{\'e}z{\'e}quel}
\affiliation{D{\'e}partement Mat{\'e}riaux Nanosciences, Institut de Physique de Rennes UMR UR1-CNRS 6251, Universit{\'e} de Rennes 1, F-35042 Rennes Cedex, France}

\author{P. Schieffer}
\altaffiliation{Corresponding author.\\ philippe.schieffer@univ-rennes1.fr}
\affiliation{D{\'e}partement Mat{\'e}riaux Nanosciences, Institut de Physique de Rennes UMR UR1-CNRS 6251, Universit{\'e} de Rennes 1, F-35042 Rennes Cedex, France}

\begin{abstract}

By combining x-ray excited Auger electron diffraction experiments and multiple scattering calculations we reveal a layer-resolved shift for the Mg $KL_{23}L_{23}$ Auger transition in MgO ultrathin films (4-6 \AA) on Ag(001). This resolution is exploited to demonstrate the possibility to control Mg atoms incorporation at the MgO/Ag(001) interface by exposing the MgO films to a Mg flux. A substantial reduction of the MgO/Ag(001) work function is observed during the exposition phase and reflects, both band-offset variations at the interface and band bending effects in the oxide film.

\end{abstract}

\date{\today}
\maketitle

Ultrathin oxide films grown on metal substrates constitute a class of materials with interesting and novel properties of their own, whose characteristics can be modified by choosing suitable metal-oxide combinations and changing the film thickness \cite{Freysoldt2007}. In particular, oxide/metal interfaces have received considerable attention in the field of catalysis \cite{Freund2008, Surnev2013}, due to their pivotal role in controlling the charging and adsorption behavior of metal nanoclusters on the oxide surface \cite{PacchioniSterrer, Benedetti2013}. To date, an important research direction involves interfaces engineering. By varying the nature of well-defined interface defects, one can tune electronic properties of the oxide/metal system such as the work function without altering the oxide overlayer \cite{Martinez2008, Martinez2009, Wlodarczyk2012, Jaouen2012}. In this context, MgO/Ag(001) represents a well-studied model system of the metal/oxide interface at the ultrathin limit \cite{Pacchioni2013}. Jung \textit{et al.} have theoretically shown that modifying the MgO/Ag(001) interface by inserting oxygen or magnesium vacancies and impurities lead to an enhanced chemical reactivity of the oxide surface with respect to the dissociation of H$_2$O \cite{Jung20112012}. Recent density functional theory (DFT) studies have further highlighted the influence of interfacial oxygen vacancies and impurities on the MgO/Ag(001) work function \cite{Ling2013, Cho2013}. Despite the increasing theoretical understanding of doped metal/oxide interfaces, few experiments dealing with post-growth interface engineering have been performed due to the practical difficulties inherent to the buried interfaces. It has been demonstrated in the case of a porous silica/Mo(112) system exposed to a Li flux, that Li atoms can penetrate the topmost silica layer and bind as Li$^+$ cations at the metal-oxide interface thereby reducing the work function of the metal/oxide system \cite{Martinez2009, Jerratsch2009}. In this paper, we demonstrate that one can successfully tune the electronic properties of a buried interface between a more compact oxide material, namely MgO, and Ag(001). By taking advantage of the nearly layer-by-layer growth mode of the MgO on Ag(001) \cite{Schintke2001, Valeri2002, Wollschlager2001}, and by combining x-ray excited Auger electron diffraction (AED) measurements and multiple scattering calculations, we evidence a layer-by-layer resolution of the Mg $KL_{23}L_{23}$ Auger transition in ultrathin MgO films (4-6 \AA) on Ag(001). We demonstrate the possibility to incorporate Mg atoms at the MgO/Ag(001) interface by simple exposures of the MgO films to a Mg flux. A gradual reduction of the metal/oxide work function upon Mg exposition (up to 0.7 eV) is observed and, in agreement with DFT calculations, is related to band-offset variations at the interface and band bending effects in the oxide film.   
  
All experiments were performed in a multi-chamber ultrahigh vacuum (UHV) system with base pressures below 2$.$10$^{-10}$ mBar. The (001)-oriented Ag single crystal was cleaned by several cycles of Ar$^+$ ion bombardment and annealing at 670-720 K. The MgO layers (4.1-6.2 \AA = 1.9-2.9 monolayer (ML)) were grown on the prepared Ag surface by evaporation of Mg in O$_2$ background atmosphere (oxygen pressure = 5$.$10$^{-7}$ mBar) at 453 K with a cube-on-cube epitaxy with respect to the Ag(001) substrate. The measurements were carried out using x-ray and ultraviolet photoelectron spectroscopy (XPS-UPS). A two axis manipulator allowed polar and azimuthal sample rotations with an accuracy better than 0.2$^{\circ}$. AED measurements were performed for the Mg $KL_{23}L_{23}$ Auger transition which leads to electrons with kinetic energies around 1177 eV and AED profiles were recorded during polar sample rotations (the polar angle is defined with respect to the surface normal) between -5$^{\circ}$ and 60$^{\circ}$ for the (100) and (110) inequivalent emission planes of the cubic structure of the MgO(001) film. The kinetic energy of the emitted electrons has been measured by employing a hemispherical analyzer (Omicron EA125) with a five-channel detection system. Al $K\alpha$ was used as the x-ray source and He-I resonance line ($h\nu=$21.22 eV) provided the UPS source for photoemission experiments. The total energy resolutions were respectively 0.80 eV and 0.15 eV for XPS and UPS.

The multiple scattering spherical wave cluster calculations have been performed in the Rehr-Albers framework \cite{Rehr1990}, by using the MsSpec program \cite{Sebilleau2006, Sebilleau2011} for clusters containing up to 420 atoms. Details on the calculations are given in Ref. \cite{Agliz1995}. Briefly, the multiple scattering expansion of the photoelectron wave function was carried out up to the fourth order which we checked to be sufficient to achieve convergence for the configurations considered. Following various experimental works \cite{Flank1996,Luches2004}, we assumed pseudomorphic ultrathin MgO films on Ag(001) with  interface Mg atoms occupying the substrate hollow sites and an interfacial distance between Ag and O atoms of 2.51 \AA. Finally, a broadening of the AED peaks due to the formation of mosaic observed during the growth of the MgO films on Ag(001) \cite{Wollschlager1998}, was taken into account by averaging the calculations over a cone of 2.5$^{\circ}$ half angle.

The DFT calculations have been carried out in the generalized gradient approximation (GGA) using the Perdew-Burke-Ernzerhof (PBE) exchange-correlation functional \cite{Perdew1996}. We have performed the calculations within the Projector-Augmented Wave (PAW) formalism \cite{Blochl1994}, implemented in a real-space grid in the GPAW code \cite{Enkovaara2010, Mortensen2005}, with a grid spacing of 0.18 \AA. The MgO/Ag (001) system was modeled with three layers of MgO deposited on three Ag layers with lattice parameter $a_0=$ 4.16 \AA~ and Ag interface atoms below the oxygen anions. $(\sqrt{2}\times\sqrt{2})a_0$ surface unit cells were used for calculations and Brillouin zone integration was performed using 4$\times$4$\times$1 Monkhorst-Pack meshes \cite{MonkhorstPack1976}. Geometry optimization was performed until the atomic forces were less than 0.02 eV/\AA. All atoms were free to move except the Ag atoms of the bottom layer. The vacuum region between adjacent slabs was set to $\sim$20 \AA.

\begin{figure}[t]
\includegraphics[width=0.48\textwidth]{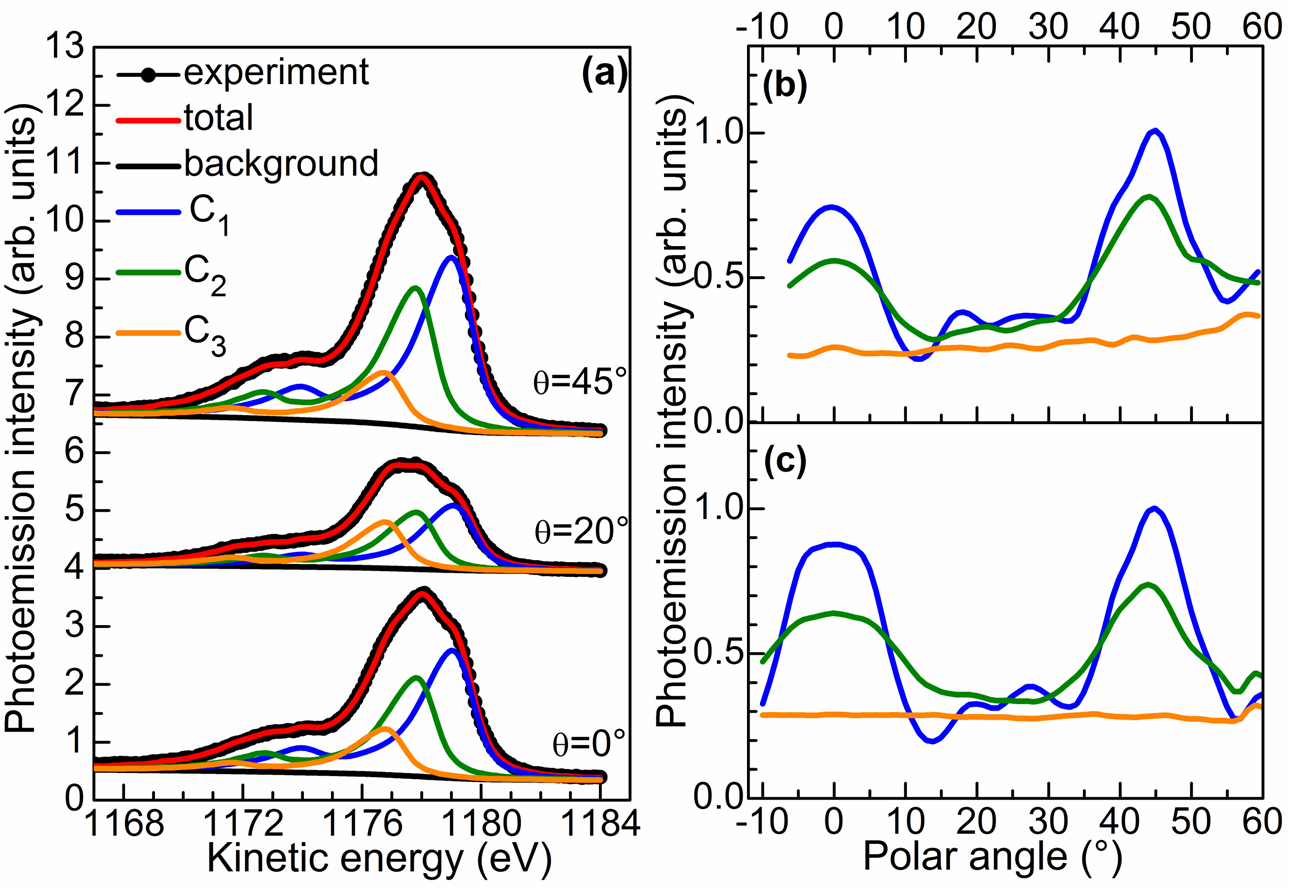}
\caption{\label{fig1} (color online). (a) Photoemission spectra of the Mg $KL_{23}L_{23}$ Auger transition of the 3 ML MgO film for polar angles of 0$^{\circ}$, 20$^{\circ}$, and 45$^{\circ}$. Best fit and layer-by-layer decomposition are also shown. (b) Experimental AED polar scans of the $C_1$, $C_2$, and $C_3$ Auger components in the (100) emission plane. (c) Calculated Mg $KL_{23}L_{23}$ AED profiles at 1177 eV in the (100) emission plane. The blue, green, and orange curves correspond respectively to calculations for Mg atom emitters in the first, second, and third layer above the MgO/Ag(001) interface.}
\end{figure}

Figure \ref{fig1}(a) shows experimental Mg $KL_{23}L_{23}$ Auger spectra of the 3 ML MgO film for polar angles of 0$^{\circ}$, 20$^{\circ}$, and 45$^{\circ}$ in the (100) emission plane. The Auger spectra (vertically shifted for clarity) exhibit unusual lineshapes which are normally constituted of two peaks separated by about 5 eV corresponding to the $^1S$ and $^1D$ multiplet structures of the Mg $2p$ final state. Depending on the polar angle, the intensities, shapes and kinetic energies of the maxima of the different spectra strongly change. The fitting procedure used an experimental Mg $KL_{23}L_{23}$ Auger spectrum taken in Ref. \cite{Altieri2000} and corresponding to a 1 ML MgO/Ag(001) sample. This spectrum was not broadened and our experimental spectra were reproduced by energy shifts of the 1 ML spectrum. The fitting procedure leads to a very good reproduction of the experimental data where the 3 ML spectra are fitted by three shifted monolayer-Auger components $C_1$, $C_2$, and $C_3$ with maxima respectively situated at 1179.1 eV, 1177.8 eV, and 1176.8 eV.
We show on Fig. \ref{fig1}(b) the experimental AED polar scans in the (100) emission plane for the three components (AED profiles are normalized with respect to the maximum value of the $C_1$ modulations). The intensity distribution of the $C_3$ component is isotropic whereas the AED profiles associated with the $C_1$ and $C_2$ contributions show pronounced peaks at normal emission and at $\sim$45$^{\circ}$ which correspond to forward-focusing peaks along the [001] and [101] atomic directions of the rocksalt (NaCl) structure of the MgO lattice. For 2 ML MgO/Ag(001), only the $C_1$ and $C_2$ components have been observed (results not shown) in the Mg $KL_{23}L_{23}$ spectrum with kinetic energies very close to those of the 3 ML sample. The analysis of the AED profiles has further revealed that for 2 ML of MgO, only monolayer and bilayer are formed on the Ag(001) surface. It can be thus expected that a $C_i$ Auger component corresponds to an Auger electron emission from the $i^{th}$ MgO layer above the metal/oxide interface. 

Our multiple scattering calculations have further confirmed this interpretation. By considering a pseudomorphic MgO film which fully covers the Ag(001) surface, we found that the $C_1$ and $C_2$ modulations in the AED profiles are caused by scattering effects on electrons emitted from Mg atoms respectively located below two and one MgO layers. Next, by comparing the experimental and calculated intensities (averaged over all polar angles in the (100) emission plane) and angular positions of the forward-focusing peaks maxima along the [010] direction, we obtained an actual MgO coverage of 2.6 ML with bilayer and trilayer proportions of $\sim$37\% and $\sim$63\%, respectively and inter-planar distance between two successive MgO planes of 2.12 \AA~ and 2.14 \AA~. These values are very close to those of 2.14-2.15 \AA~ determined by Luches \textit{et al.} by using Extended X-ray absorption fine structure (EXAFS) on a 3 ML MgO/Ag(001) sample \cite{Luches2004}. As it can be seen in Fig. \ref{fig1}(c), the simulated AED profiles integrating the morphological parameters, are in remarkable agreement with the experimental AED profiles shown in Fig. \ref{fig1}(b). All these findings indicate that we obtain a straightforward experimental evidence of a layer-by-layer resolution of the Mg $KL_{23}L_{23}$ Auger transition for MgO ultrathin films on Ag(001). 
\begin{figure}[b]
\includegraphics[width=0.48\textwidth]{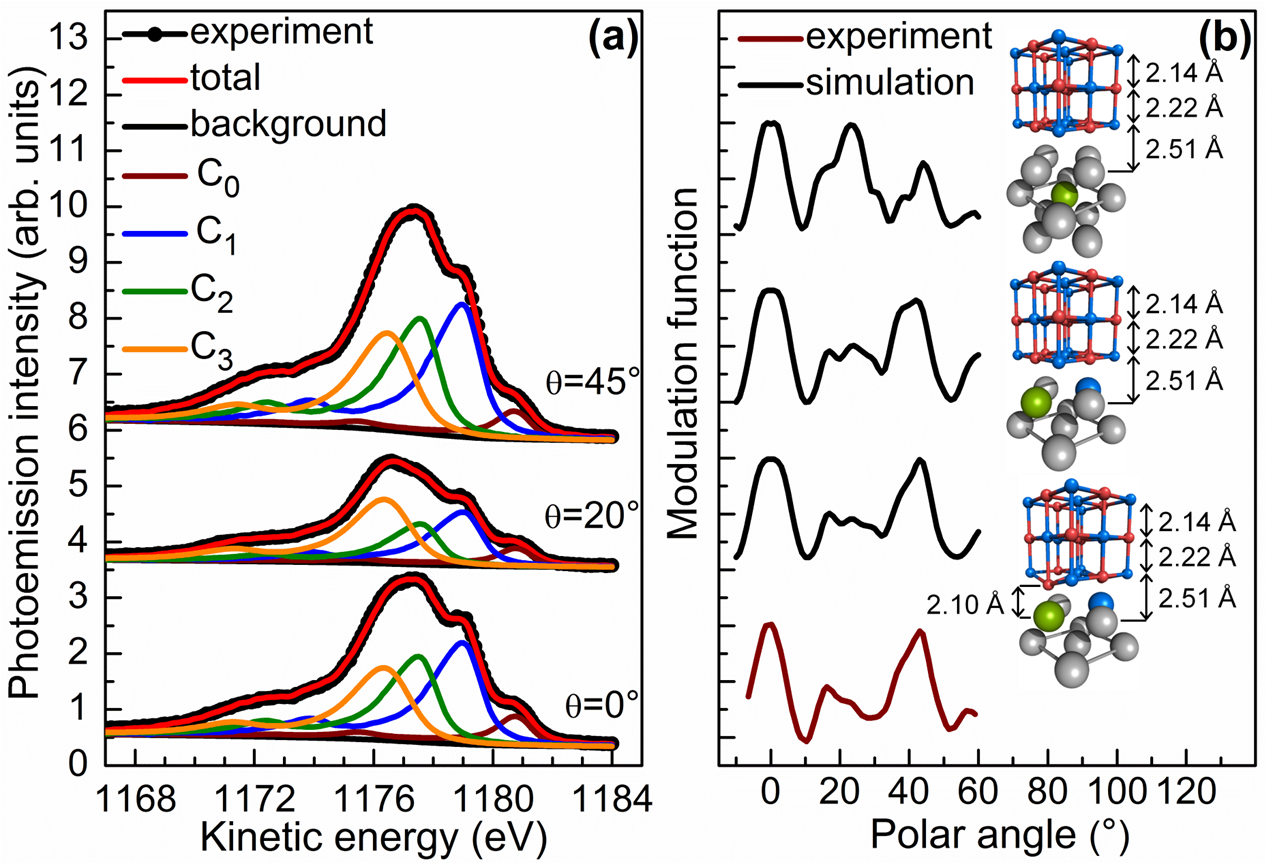}
\caption{\label{fig2} (color online). (a) Photoemission spectra of the Mg $KL_{23}L_{23}$ Auger transition of the 3 ML MgO film for polar angles of 0$^{\circ}$, 20$^{\circ}$, and 45$^{\circ}$ after exposition to an Mg atomic flux. Best fit and layer-by-layer decomposition are also shown. (b) Comparison between experimental (bottom curve) and calculated modulation functions associated with the $C_0$ component along the [100] direction for different configurations with Mg atoms occupying substitutional sites of the Ag substrate. The basic cells used for the multiple scattering calculations are reported on the right side. The red, grey and blue atoms correspond respectively to oxygen, silver and Mg scatters. The incorporated Mg emitter atom is also shown in green.}
\end{figure}

Similar distance-dependent relaxation shifts of photoemission and Auger energies have been observed by Kaindl \textit{et al.} for Xe multilayers on Pd(001) \cite{Kaindl1980}, and have been related to the presence of an image potential screening. More recently, this behavior has been also evidenced in the case of MgO/Ag(001) by Altieri \textit{et al.} \cite{Altieri1999}. Note, however, that the layer-by-layer resolution of the Mg $KL_{23}L_{23}$ transition has never been identified before. We further believe that these layer-resolved Auger shift could be observed for other ultrathin oxide layers deposited on highly polarizable medium such as metals and have therefore to be considered to obtain quantitative values for physical parameters such as Coulomb or charge-transfer energies.

We now use this layer-by-layer resolution to demonstrate the Mg atom incorporation at the MgO/Ag(001) interface. Figure \ref{fig2}(a) shows the Mg $KL_{23}L_{23}$ Auger spectra of a 3 ML MgO/Ag(001) sample after exposition to an Mg atomic flux (2.4$\times$10$^{13}$ atoms/(cm$^{2}$s)) during 12 minutes at a substrate temperature of 513 K to avoid the formation of Mg metallic clusters. In addition to the $C_1$, $C_2$, and $C_3$ components, a fourth component (labeled $C_0$) at higher kinetic energy (1180.0 eV) is needed to obtain satisfactory fits of the experimental spectra. This component appears at a kinetic energy (1180.2 eV) very close to that obtained for submonolayer depositions of Mg on Ag(001) and is actually a typical spectroscopic fingerprint of electron emission from Mg atoms located in a metallic environment.
 
Figure \ref{fig2}(b) shows, in turn, the modulation function associated with the $C_0$ component along the [100] direction (the modulation function corresponds to normalized experimental or calculated AED curves forced to have an amplitude between -0.5 and 0.5). The modulations reveal a well-structured AED pattern similar to those of the $C_1$, $C_2$ components with forward-focusing peaks along the [001] and [101] atomic directions. Knowing that the $C_0$ component corresponds to electron emission from Mg atoms located in a metallic environment, we can therefore expect that Mg atoms substitute Ag atoms in the vicinity of the MgO/Ag(001) interface. Hence, we have calculated modulation functions for different configurations with Mg atoms occupying substitutional sites of the Ag substrate (the basic cells used for the multiple scattering calculations are reported on the right side of Fig. \ref{fig2}(b)). The best agreement with the experimental profile is obtained for a simulation with an Mg emitter atom located in the first Ag layer, beyond the 3 ML of the MgO film (third simulation curve at the bottom of Fig. \ref{fig2}(b)) and a quantitative analysis of the AED data indicates that the Mg atom composition in the first substrate layer is about 30\%. In other words, we demonstrate the remarkable fact that Mg atoms can be incorporated at MgO/Ag(001) interface by simple Mg flux exposures of the MgO film. As it can be seen in Fig. \ref{fig2}(b), a distortion of the interface layers leads to a better agreement between experimental and simulated data. Such a distortion upon Mg incorporation has been further predicted by our DFT calculations. For an incorporation of 50 \% of Mg atoms, our DFT analysis has shown that only the MgO interface layer undergoes a significant distortion with the O$^{2-}$ ions located above the Ag (Mg) metal atoms displaced outwards (downwards) by 0.14 \AA~ (0.16 \AA) relative to the Mg$^{2+}$ ions position. In addition, a slight atomic corrugation in the interfacial alloy has been predicted with the incorporated Mg atoms displaced towards the MgO lattice by 0.18 \AA. 

\begin{figure}[t]
\includegraphics[width=0.48\textwidth]{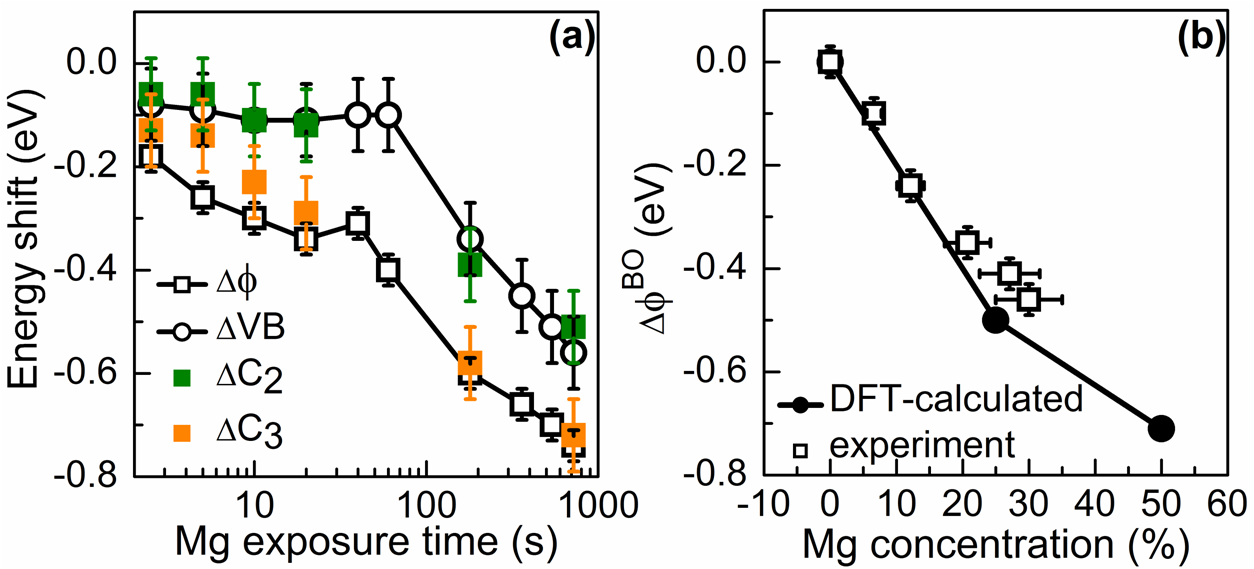}
\caption{\label{fig3} (color online). (a) Variations of the work function $\Delta\phi$, MgO-valence band position $\Delta$VB and kinetic energies of the $C_2$ ($\Delta C_2$) and $C_3$ ($\Delta C_3$) Auger components as a function of the Mg exposure time for a 3 ML MgO/Ag(001) sample. The evolutions are plotted in semi-logarithmic representation. (b) Comparison between DFT-calculated and experimental work function changes $\Delta\phi^{BO}$ as a function of the Mg concentration at the metal/oxide interface. $\Delta\phi^{BO}$ reflects the band-offset (BO) variations at the metal/oxide interface.}
\end{figure}

In Fig. \ref{fig3}(a) are shown the variations of the work function ($\Delta\phi$), MgO-valence band position ($\Delta$VB) and kinetic energies of the $C_2$ and $C_3$ components (respectively labeled $\Delta C_2$ and $\Delta C_3$) as a function of the Mg exposition time for a 3 ML MgO/Ag(001) sample. The methods used for the determination of the work function and valence band (VB) position from our UPS spectra are given in Ref. \cite{Jaouen2010}. A semi-logarithmic representation has been chosen to highlight the two-step evolution of the considered quantities. During the first 20 seconds, we observe that $\Delta$VB and $\Delta C_2$ follow the same evolution and that $\Delta C_3$ is about two times higher than $\Delta C_2$. Thus, the $\Delta C_2$ changes reflect half of a total downward band bending of $\sim$0.25-0.30 eV after 20 seconds. In meantime the work function changes by $\sim$-0.35 eV suggesting that only a small additional dipole is created at the oxide surface (leading to a work function variation lower than -0.1 eV).
As shown by first-principle calculation \cite{Geneste2005}, single Mg atom is weakly bonded to oxygen on flat MgO(001) surface and can easily desorb at 513 K. In contrast, it binds rather strongly to the low-coordinated sites of MgO defective surfaces \cite{Kantorovich1999}. Electron paramagnetic resonance (EPR) experiments have further shown that the deposition of a very small quantity of Mg atoms on an MgO(001) film at 50 K leads to the formation of positively charged color centers (anion vacancies) at morphological defects of the MgO surface \cite{Gonchar2010}. These results have been recently supported by a DFT study of the MgO/Ag(001) system \cite{Ling2013}, in which it has been found that oxygen vacancies with two and one trapped electrons (respectively labeled $F^0$ and $F^+$ centers) can coexist at corner and edge sites of the rough MgO surface. Whereas the introduction of an $F^0$ neutral surface center does not modify the work function of the metal/dielectric system, the appearance of positively charged $F^+$ centers leads to its substantial reduction. For instance, a work function decrease of 0.37 eV was obtained for 1.6$\times$10$^{13}$ defects/cm$^2$. We thus conclude that the band bending effect and associated work function decrease observed in the initial phase of the Mg exposure results from the positive charge accumulation in surface color centers.
 
Finally, after 40 seconds, similar changes of $\Delta\phi$, $\Delta$VB, $\Delta C_2$ and $\Delta C_3$ ($\sim$-0.40 eV between 40 and 720 seconds) are observed in Fig. \ref{fig3}(a). This demonstrates that a progressive change of the Fermi level pinning position at the MgO/Ag(001) interface takes place and that the band bending is quasi-constant during this second phase. Figure \ref{fig3}(b) shows the comparison between our DFT-calculated and experimental work function changes of the second phase as a function of the Mg concentration in the Ag interface layer. The concentrations have been estimated from the intensities of the $C_0$ Auger component by comparing them with that measured for an exposure time of 720 s for which we have obtained an Mg composition of 30 \% from the AED analysis. As it can be seen, the incorporation of Mg atoms at the interface leads to a strong work function reduction of the metal/oxide system. The experimental $\Delta\phi^{BO}$ values are in overall good agreement with the DFT-predicted work function changes and essentially reflect the band-offset (BO) variation at the metal/oxide interface which is mainly correlated to the increase of the electrostatic compression effect due to the presence of interfacial Mg atoms \cite{Jaouen2012, Goniakowski2004}.  

In conclusion, by taking advantage of the layer-by-layer resolution of the Mg $KL_{23}L_{23}$ Auger emission for MgO ultrathin films on Ag(001) we have demonstrated that Mg atoms can be incorporated at the MgO/Ag(001) interface by simple exposures of the MgO film to a Mg flux. Our experiments have also shown the possibility to probe the evolution of the dipoles through and at the surface of an ultrathin oxide layer and have thus allowed a precise identification of the mechanisms responsible for the work function decrease upon Mg exposure. We have found that these reductions are related to Fermi-level pinning modification at the interface and band bending effects in the oxide film due to the formation of charged color center at the oxide surface.

\begin{acknowledgments}
The authors warmly acknowledge A. Le Pottier and Y. Claveau for technical support. Parts of this work have been funded by European FP7 MSNano network under Grant agreement n$^{\circ}$ PIRSES-GA-2012-317554 and partially supported by the Fonds National Suisse pour la Recherche Scientifique through Division II and the Swiss National Center of Competence in Research MaNEP.
\end{acknowledgments}



\end{document}